\documentstyle[osa,manuscript,psfig]{revtex}
\begin{document}

\titlepage
\title{
Hyperon polarization in semi-inclusive deeply inelastic
lepton-nucleon scattering at high energy} 
\author {Liu Chun-xiu, Xu Qing-hua, and Liang Zuo-tang}
\address {Department of Physics,
Shandong University, Jinan, Shandong 250100,China}

\maketitle

\begin{abstract}     
We calculate the polarizations
for different octet hyperons 
produced in the current fragmentation regions 
of the deeply inelastic lepton-nucleon scatterings
$\mu^-N \to \mu^- HX$ and
$\nu_{\mu} N \to \mu^- HX$ at high energy
using different models for spin transfer 
in fragmentation processes.
The results show that measurements 
of those hyperon polarizations
should provide useful information 
to distinguish between different models
in particular the SU(6) 
and the DIS pictures used frequently in the literature.
We found, in particular, that
measuring the polarization of $\Sigma^+$  produced 
in these processes can give a better test 
to the validity of the different spin transfer models.
\end{abstract}     

\newpage

\section{Introduction}
Spin transfer in high-energy hadronization process
of the fragmenting quark to the produced hadron
is one of the important issues
in the spin effects in hadronization processes.
The problem has attracted 
much attention [\ref{att}-\ref{Ma0002}] recently.
This is partly triggered by the ALEPH and OPAL
measurements\cite{ALEPH96,OPAL98} on $\Lambda$ polarization 
in $e^+e^-$ annihilation at the $Z^0$ pole. 
Compared with the theoretical calculations\cite{GH93,BL98},
those data\cite{ALEPH96,OPAL98}
seem to suggest that 
the simple SU(6) wave-function 
can be used to describe 
the relation between the spin of the fragmenting quark
and that of the produced hyperon 
which contains the fragmenting quark.
This is rather surprising 
because the energy
is very high at CERN $e^+e^-$ collider LEP and the initial
quarks and antiquarks produced
at the annihilation vertices of
the initial $e^+e^-$ are certainly
current quarks and current antiquarks
rather than the constituent quarks used
in describing the static properties of hadrons
using SU(6) symmetric wave-functions.
A conclusive judgment of  
the different spin transfer models
can still not be made.
It is thus interesting and instructive to
make further checks in experiments by making
complementary measurements.
For this purpose, we have made a systematic study
of hyperon polarizations in different lepton-induced
reactions using the SU(6) or the DIS picture.
The results for $e^+e^-$ annihilation
have been presented in Ref. [\ref{LL2000}].
We now present the results for 
other lepton-induced reactions,
such as $\mu^-N \to \mu^- HX$ and
$\nu_{\mu} N \to \mu^- HX$.

Compared with those in $e^+e^-$ annihilations,
hyperon polarizations in the current fragmentation regions
of the deeply inelastic lepton-nucleon scatterings
$\mu^-N \to \mu^- HX$ and
$\nu_{\mu} N \to \mu^- HX$
at high energy
are of particular interests for the following reasons:
First, not only longitudinally but also transversely
polarized quark beam can be produced 
in the current fragmentation regions
of $\mu^-N \to \mu^- HX$.
We can study here not only longitudinal transfer
but also check 
whether it is the same for the transverse polarization case.
Second, there is an automatic flavor separation 
in some cases in the deeply inelastic lepton-nucleon scatterings.
For example, in the current fragmentation region of
$\nu_{\mu} N \to \mu^- HX$, 
we have a predominant contribution from the $u$ quark fragmentation.
In $\mu^- N \to \mu^- HX$,
we have a combined contribution 
from the $u$, $d$, and $s$-quarks 
with the corresponding weights ${4\over9}u(x)$,
${1\over9}d(x)$, and ${1\over9}s(x)$,
where $x$ is the Bjorken-$x$, and
$q(x)$ is the number density of quark in proton.
Since the $x$ dependences of $u(x)$, $d(x)$ and $s(x)$
are different from each other,
we can tune the contributions of 
different flavors by choosing events in different $x$-region.
Since different flavors contribute quite differently 
to the polarizations of different hyperons,
we expect that the results obtained 
in different kinematic regions 
should be quite different from each other and 
they should also be different from those obtained 
in $e^+e^-$ annihilations\cite{LL2000}. 
Measurements of these polarizations with high accuracy  
should provide us with detailed tests of 
different aspects of different models.

The paper is arranged as follows: 
In Sec. 2, 
we briefly summarize the calculation method 
of the hyperon polarization, which was  
outlined in Ref. [\ref{LL2000}] using $e^+e^-\to \Lambda X$ 
as an example, and discuss the differences 
in different reactions.
In Sec. 3, 
we present the results for the hyperon polarizations 
in the current fragmentation regions 
of the deeply inelastic scatterings off nucleon
with $\mu^-$ or $e^-$ beam.
In Sec. 4, we present
our results for the hyperon polarization 
in the current fragmentation region of
$\nu_{\mu} N \to \mu^- HX$.
A brief summary of the results is given in Sec. 5.

\section{The calculation method}
\label{section1}

The method of calculating
the longitudinal polarization $P_{H_i}$ of different
hyperon  $H_i$ in the fragmentation of a longitudinally 
polarized quark $q^0_f$ 
has been outlined in Ref. [\ref{LL2000}] using 
the inclusive process $e^+e^-\to  H_i+X$ as an example.
The method is independent of the way how 
$q_f^0$ is produced thus 
can certainly be applied to other processes
where the longitudinally polarized 
fragmenting quarks are produced.
Similarly, it can also be applied to processes 
where transversely polarized quarks are produced.
We now briefly summarize the main points 
and discuss the differences in different reactions 
in the following.

We recall that, 
to calculate the polarization of hyperon $H_i$'s which are produced
in the fragmentation of a polarized quark $q^0_f$,
we should consider the $H_i$'s 
which have the following different origins separately. 

(a) Hyperons which are directly produced 
and contain the fragmenting quark $q^0_f$;

(b) Hyperons which are decay products of other heavier 
hyperons which were polarized before their decay; 

(c) Hyperons which are directly produced but 
do not contain the fragmenting quark $q^0_f$;

(d) Hyperons which are decay products of other heavier hyperons 
which were unpolarized before their decay. 

It is clear that hyperons 
from (a) and (b) can be polarized
while those from (c) and (d) are not \cite{GH93,BL98,LL2000}.  
Hence,  
the polarization of $H_i$ is given by
\begin{equation}
P_{H_i}={ {\sum\limits_f t^F_{H_i,f} P^{(q)}_f \langle n^a_{H_i,f}\rangle
+\sum\limits_{j} t^D_{H_i, H_j} P_{H_j} \langle n^b_{H_i, H_j}\rangle}
 \over
{\langle n^a_{H_i}\rangle +\langle n^b_{H_i}\rangle + 
\langle n^c_{H_i}\rangle +\langle n^d_{H_i}\rangle} }. 
\label{eq1}
\end{equation}
Here, $P^{(q)}_f$ is the polarization of the fragmenting quark $q^0_f$;
$\langle n^a_{H_i,f}\rangle$ is the average number of
the hyperons which are directly produced and contain
the fragmenting quark of flavor $f$;
$\langle n^b_{H_i,H_j}\rangle$ is the average number of $H_i$
coming from the decay of $H_j$,
$P_{H_j}$ is the polarization of $H_j$
before its decay;
$\langle n^a_{H_i}\rangle(\equiv \sum\limits_f \langle n^a_{H_i,f}\rangle$),
$\langle n^b_{H_i}\rangle(\equiv \sum\limits_j \langle n^b_{H_i,H_j}\rangle)$,
$\langle n^c_{H_i}\rangle$, and $\langle n^d_{H_i}\rangle$
are average numbers of hyperons in group (a), (b), (c), 
and (d), respectively;
$t^F_{H_i,f}$ is the ``fragmentation polarization transfer factor"
from quark $q^0_f$ to hyperon $H_i$,
where the superscript $F$ stands for fragmentation;
$t^D_{H_i,H_j}$ is the ``decay polarization transfer factor"
from $H_j$ to $H_i$ in the decay process
$H_j \to H_i+X$,
where the superscript $D$ stands for decay.

We recall that the fragmentation polarization
transfer factor $t^F_{H_i,f}$ is equal to the fraction of
spin carried by the $f$-flavor-quark
divided by the average number of quark of flavor $f$
in the hyperon $H_i$\cite{GH93,BL98,LL2000}.
It is different in the SU(6) or the DIS picture.
The decay polarization transfer factor $t^D_{H_i,H_j}$
is determined by the decay process.
Clearly, both of them are universal 
in the sense that they are independent
of the processes in which $q^0_f$ and $H_j$ are produced.
They are the same in $e^+e^-$ annihilations
and in the deeply inelastic scattering with longitudinally polarized
beam and/or target and in that with neutrino beam.
These two factors are discussed in detail in Ref. [\ref{LL2000}].
They can be found, e.g., 
in table I and table II in Ref. [\ref{LL2000}], respectively.

The differences for different processes
come from the polarization $P^{(q)}_f$ of 
the fragmenting quark $q^0_f$ 
and the average numbers $\langle n^a_{H_i,f}\rangle$,
$\langle n^b_{H_i,H_j}\rangle$, $\langle n^c_{H_i}\rangle$,
and $\langle n^d_{H_i}\rangle$ for hyperon $H_i$
of the different origins. 
Since the leading order hard subprocesses 
of these lepton-induced reactions
involve only electro-weak interactions,
the polarization $P^{(q)}_f$ of the fragmenting quark $q_f^0$ 
produced in these reactions
can easily be calculated using the standard model of
electro-weak interactions.
They are in general different for $e^+e^-$ annihilation 
and for deeply inelastic lepton-nucleon scatterings. 
They are also different in different kinematic regions 
in deep-inelastic scattering or for different flavors 
of the struck quarks. 
Since different flavors contribute quite differently 
to the polarizations of different hyperons, 
this makes the situation full of abundance and interesting.    
We will discuss them separately in next sections.

In the fragmentation process of a 
given flavor of quark $q_f^0$ at a given energy, 
the average numbers $\langle n^a_{H_i,f}\rangle$,
$\langle n^b_{H_i,H_j}\rangle$, 
$\langle n^c_{H_i}\rangle$,
and $\langle n^d_{H_i}\rangle$ 
should also be independent of the process 
in which $q_f^0$ is produced. 
They are determined by the hadronization mechanism 
and should be independent of the polarization of $q_f^0$.
But, for different reactions, 
the relative weights of the contributions of 
different flavors are different. 
Hence the resulting  $\langle n^a_{H_i,f}\rangle$,
$\langle n^b_{H_i,H_j}\rangle$, 
$\langle n^c_{H_i}\rangle$,
and $\langle n^d_{H_i}\rangle$ are also different 
in different reactions. 
For deeply inelastic lepton-nucleon scatterings, 
the relative weights are determined by the 
number densities of different flavors in the nucleon. 
This implies that the average numbers 
$\langle n^a_{H_i,f}\rangle$,
$\langle n^b_{H_i,H_j}\rangle$, 
$\langle n^c_{H_i}\rangle$,
and $\langle n^d_{H_i}\rangle$ are 
determined by the hadronization mechanism 
and the structure functions of the nucleon.
We emphasize here that, 
although neither the structure function 
nor the hadronization mechanism 
is theoretically clear yet, 
the form of the unpolarized parton densities in nucleon 
and that of the unpolarized fragmentation functions 
are empirically known to reasonably high accuracy. 
We can calculate them using the parameterizations 
of the structure functions and the phenomenological 
hadronization models. 
Since we are not concerned with the 
different correlations among the produced particles, 
different hadronization models lead essentially to the same results. 
Presently, these results can be obtained 
conveniently from the event generators 
using Monte Carlo method. 
We used the Lund string fragmentation model\cite{AGIS83}
implemented by {\scriptsize LEPTO}\cite{LEPTO} in our calculations
for deeply inelastic lepton-nucleon scattering.

\section{Hyperon polarization in the current fragmentation region of 
$\mu^- N \to \mu^- HX$}

For sufficiently high $Q^2$ and high total hadronic energy $W$, 
hadrons produced in the current fragmentation region 
in deeply inelastic lepton-nucleon scattering 
can be considered as a pure result of the fragmentation 
of the single struck quark. 
(Here, $Q^2$ is the virtuality of the exchanged 
boson in the initial electro-weak interaction; 
$W^2=Q^2(1/x-1)+M^2$, and $M$ is the nucleon mass). 
Hence, they can be used to study 
the spin effects in high-energy fragmentation processes, 
in particular the spin transfer from the polarized 
fragmenting quark to the produced hadrons. 
We will limit ourselves to such kinematic regions 
in this paper and will not 
discuss the cases at lower energies\cite{HERMES,NOMAD} where 
no distinct separation between 
the fragmentation results of the struck quark 
and that of the rest of the target nucleon is possible\cite{ref1}.  
We will discuss both the case where 
the virtual photon exchange dominates and  
the case where charged weak current 
interaction takes place.  
In this section, we will discuss the first case using 
$\mu^-N \to \mu^- HX$ as an example.
The latter case will be discussed in Sec. 4. 

In the deeply inelastic scattering $\mu^-N \to \mu^- HX$,
the leading order hard subprocess $\mu^-q\to\mu^-q$
is an electro-magnetic 
process in which a single virtual photon exchange dominates.
In such subprocess,
if the initial $\mu^-$ or $q$ is polarized 
or both of them are polarized,
the outgoing quark can be polarized.
Obviously, the polarization of the outgoing 
quark is dependent on  
the polarization of the initial $\mu^-$, 
that of $q$, and the scattering process.
There should be a factor relating the 
polarization of the outgoing quark with that 
of the initial quark or $\mu^-$. 
This factor is usually referred as 
the depolarization factor. 
In the following, we will first briefly 
summarize the depolarization factor 
for the elementary subprocess $\mu^-q\to\mu^-q$, 
and then present the results that we obtained 
for hyperon polarization in the current 
fragmentation region of $\mu^-N \to \mu^- HX$
using different pictures for spin transfer 
in the fragmentation process.
We will also briefly discuss 
how to tune the contributions of
different flavors by choosing different $x$ region.

\subsection{Depolarization factor in $\mu^-q\to\mu^-q$}
\label{sub1}

The calculation of the polarization of the outgoing 
quark from $\mu^-q\to\mu^-q$ is straightforward 
and the results can be found in different publications. 
For completeness, we briefly summarize the main 
ingredients used in the calculations and 
the main results which will be used in our calculations 
of hyperon polarization and discuss 
the behaviors of these results in 
different cases in this subsection. 
   
We recall that the polarization of a 
particle system is described by 
the spin density matrix $\rho$. 
For the system of the incoming $\mu^-$ and $q$, 
which are independent of each other, 
$\rho$ is a direct product of 
that of $q$ and that of $\mu^-$, 
which we denote by $\rho^{(q)}$ 
and $\rho^{(l)}$ respectively.
The $\rho^{(q)}$ and $\rho^{(l)}$ 
are $2\times 2$ matrices, which is taken as the form 
$\rho^{(x)}=(1+\vec \sigma \cdot \vec P^{(x)})/2$; 
where $\vec \sigma$ is the Pauli matrix and 
$\vec P^{(x)}$ is the polarization vector of $x(=q$ or $l$).   
If we know the scattering matrix $M$ of 
the process, we can calculate  
the spin density matrix $\rho^{out}$ for the 
outgoing system from the $\rho^{in}$ for the 
initial system using the relation 
$\rho^{out}=M\rho^{in}M^\dag$, or in helicity bases
\begin{equation}
\rho^{out}_{\lambda'_1 h_1',\lambda'_2 h'_2}=
\sum_{\lambda_1 h_1,\lambda_2 h_2} 
M_{\lambda'_1 h'_1,\lambda_1 h_1} 
\rho^{in}_{\lambda_1 h_1,\lambda_2 h_2} 
M^*_{\lambda'_2 h'_2,\lambda_2 h_2},
\label{sub1eq1} 
\end{equation}
where we use $\lambda$ and $h$ to denote 
the helicities of $\mu^-$ and $q$ respectively, 
and the prime to denote final state; 
$M_{\lambda' h',\lambda h}$ is the usual 
helicity amplitude for the scattering process 
$\mu^-q\to\mu^-q$. 
We neglect the $\mu^-$ and $q$ mass, 
so that helicity is conserved, i.e., 
$M_{\lambda' h',\lambda h}\not=0$ only if 
$h'=h$ and $\lambda'=\lambda$. 
Furthermore, parity conservation requires
$M_{++,++}=M_{--,--}$ and $M_{+-,+-}=M_{-+,-+}$.  
Hence, there are only two nonzero independent 
$M_{\lambda' h',\lambda h}$'s, and they are
\begin{equation}
M_{++,++}=-4e^2e_q/(1-\cos\theta),
\end{equation}
\begin{equation}
M_{+-,+-}=-2e^2e_q(1+\cos\theta)/(1-\cos\theta),
\label{sub1eq2}
\end{equation}
where $\theta$ is the scattering angle in 
the $\mu^-q$ center of mass frame.
We insert these $M_{\lambda' h',\lambda h}$'s 
into Eq. (\ref{sub1eq1}), sum over the helicity 
of the outgoing $\mu^-$, and we obtain
the spin density matrix for the outgoing quark as 
\begin{equation}
\rho^{(q)out}=\left (
  \begin{array}{cc}
   C_1\rho^{(q)in}_{++}   & C_3\rho^{(q)in}_{+-} \\ 
   C_3^*\rho^{(q)in}_{-+} & C_2\rho^{(q)in}_{--} \\
  \end{array}
  \right ). 
\label{sub1eq3}
\end{equation}
Here $C_1=1+D_L(y)P^{(l)}_{L}$;
     $C_2=1-D_L(y)P^{(l)}_{L}$;
     $C_3=D_T(y)$;
where $L$ means the component parallel to the moving direction; 
$y\equiv p\cdot (k-k')/p\cdot k =(1-\cos\theta)/2$,  
where $p$, $k$, and $k'$ are the four momenta of the incoming 
$q$, $\mu^-$, and the outgoing $\mu^-$ respectively; and  
\begin{equation}
D_L(y)=[1-(1-y)^2]/[1+(1-y)^2];
\label{sub1eq4}
\end{equation}
\begin{equation}
D_T(y)=2(1-y)/[1+(1-y)^2].
\label{sub1eq5}
\end{equation}
In obtaining these results from Eqs. (\ref{sub1eq1})-(\ref{sub1eq2}), 
we have dropped a 
common factor $2e^4e_q^2[1+(1-y)^2]/y^2$, which 
has no influence on the results of 
the polarization of the outgoing quark. 

From these results, we 
can calculate the polarization of 
the outgoing $q$ in different cases. 
We first discuss the case in which 
the initial quark is not transversely polarized, 
i.e., $\rho^{(q)in}_{+-}=\rho^{(q)in}_{-+}=0$. 
In this case, we have  
$\rho^{(q)out}_{+-}=\rho^{(q)out}_{-+}=0$. 
This means that, independent of the polarization 
of the incoming lepton $\mu^-$,  
the outgoing quark is not transversely polarized. 
But, from Eq. (\ref{sub1eq3}), we see also that, in this case,  
the outgoing quark can be longitudinally 
polarized and the polarization is given by 
\begin{equation}
P_L^{(q)out}={D_L(y)P^{(l)}_{L}+P_L^{(q)in}
\over  1+P^{(l)}_LP^{(q)in}_LD_L(y)},
\label{sub1eq6}
\end{equation}
where $P_L^{(q)in}=\rho^{(q)in}_{++}-\rho^{(q)in}_{--}$ 
is the longitudinal polarization of the incoming quark. 
We see clearly that, not only the longitudinal 
polarization of the incoming $q$ but also 
the longitudinal polarization of 
the incoming $\mu^-$ can be transfered 
to the outgoing $q$. 
Furthermore, from Eq. (\ref{sub1eq6}), we see that,
\begin{equation}
 P_L^{(q)out}=P_L^{(q)in}, {\rm \ for \ \ } P^{(l)}_L=0,
\label{sub1eq7}
\end{equation} 
which shows that the 
longitudinal polarization of the quark will 
be completely transferred to the outgoing quark, 
which is just the result of helicity conservation.
We also see from Eq. (\ref{sub1eq6}) that,
\begin{equation}
 P_L^{(q)out}=P^{(l)}_LD_L(y), {\rm \ for \ \ } P_{L}^{(q)in}=0.
\label{sub1eq8}
\end{equation} 
It shows that, 
for the scattering of a longitudinally polarized lepton
on unpolarized nucleon target, 
the polarization of the incoming lepton can also be 
transferred to the outgoing quark with a reduction 
factor $D_L(y)$.  
Hence, $D_L(y)$ is usually referred to as the 
longitudinal depolarization factor. 
It is interesting to see that 
$D_L(y)$ is only a function of $y$. 
Its $y$ dependence is shown by the solid line in Fig.1.
We see that $D_L(y)>0$ for $0<y<1$,  
it increases with the increasing $y$
and $D_L(y=0)=0$, $D_L(y=1)=1$.
We recall that, for given energy and $Q^2$, 
$y\propto 1/x$. 
This implies that the depolarization factor 
$D_L$ is large for small $x$ but small for large $x$.

We denote the longitudinal polarization of 
the incoming nucleon by $P^{(N)}_L$. 
At a given Bjorken-$x$, 
for quark of flavor $f$, 
$P^{(q)in}_{fL}=P^{(N)}_L \Delta q_f(x)/q_f(x)$;
where $q_f(x)$ and $\Delta q_f(x)$ 
are the unpolarized 
and longitudinally polarized distribution functions 
of quark of flavor $f$ inside the nucleon.
Hence, we obtain the polarization of the outgoing 
quark of flavor $f$ as 
\begin{equation}
P_{fL}^{(q)out}=\frac{P^{(l)}_L D_L(y) q_f(x)+P^{(N)}_L \Delta q_f(x)}
{q_f(x) +P^{(l)}_L D_L(y) P^{(N)}_L \Delta q_f(x)}.
\label{sub1eq9}
\end{equation}
This is the general formula 
which can be used to replace the $P^{(q)}_f$ in Eq. (\ref{eq1}) 
to calculate $P_{Hi}$.
To see that $x$ dependence of $P_{fL}^{(q)out}$ in 
different cases, 
we show the numerical results obtained 
for different $P^{(l)}_L$ and $P^{(N)}_L$ 
at $Q^2=10$ GeV$^2$ and $E_{\mu}$=$500$GeV 
using the GRSV parameterization \cite{GRSV2000} 
for parton distributions in Fig.2. 
From these results, we see clearly that 
the $P_{fL}^{(q)out}$ are different for the different flavors,
and there are considerably large differences
between the results for  
different $P^{(l)}_L$ and $P^{(N)}_L$.

Now, we discuss the second case, 
in which the incoming quark is transversely polarized.
In this case, we have $\rho^{(q)in}_{++}=\rho^{(q)in}_{--}=1/2$, 
and $\rho^{(q)in}_{+-}=\rho^{(q)in*}_{-+}\not=0$. 
This means that the outgoing quark is transversely polarized. 
Furthermore, since $C_3$ is a real number, 
the direction of the transverse polarization 
with respect to the helicity frame remains the same. 
The magnitude of the transverse polarization
of the outgoing quark 
is given by 
\begin{equation}
P_{T}^{(q)out}=D_T(y)P_T^{(q)in},
\label{sub1eq10}
\end{equation}
which is independent of 
the polarization of the incoming lepton $\mu^-$.
The function $D_T(y)$, 
which is given by Eq. (\ref{sub1eq5}), is called 
transverse depolarization factor. 
It depends also only on $y$ and 
its $y$ dependence is given by the 
dashed line in Fig.1.
We see that, in contrast to $D_L(y)$, 
$D_T(y)$ decreases with the increasing $y$
and $D_T(y=0)=1$, $D_T(y=1)=0$. 

Suppose the target nucleon is transversely polarized 
with polarization $P^{(N)}_T$, at a given $x$, 
for the incoming quark of flavor $f$, 
$P_{fT}^{(q)in}=P^{(N)}_T\delta q_f(x)/q_f(x)$, 
where $\delta q_f(x)$ is the transversely polarized 
distribution function of quark of flavor $f$.
Hence, the polarization of 
the outgoing quark of flavor $f$
is given by 
\begin{equation}
P_{fT}^{(q)out}=
P^{(N)}_T \frac{\delta q_f(x)}{q_f(x)}D_T(y).
\label{sub1eq11}
\end{equation}
This should be used to replace the $P^{(q)}_f$ in Eq. (\ref{eq1}) 
to calculate the hyperon polarization in transversely 
polarized case. 

\subsection{Hyperon polarization in the current 
fragmentation region in $\mu^-N\to\mu^-HX$} 

Having the $P^{(q)}_f$'s obtained in the last subsection, 
we can calculate $P_{Hi}$ using Eq. (\ref{eq1}) if we know 
the average numbers $\langle n^a_{H_i,f}\rangle$,
$\langle n^b_{H_i,H_j}\rangle$,
$\langle n^c_{H_i}\rangle$,
and $\langle n^d_{H_i}\rangle$. 
We will present the results in different cases in 
this subsection.

\subsubsection{$\Lambda$ polarization in $\mu^-N\to\mu^-\Lambda X$}

Among all the $J^P={1\over 2}^+$ hyperons, 
$\Lambda$ is most copiously produced, and 
it can also be measured easily through 
the decay channel $\Lambda \to p\pi^-$.  
Hence, till now, most of the studies
[\ref{att}-\ref{OPAL98},\ref{NOMAD}-\ref{E665}] 
are concentrated on $\Lambda$ polarization. 
However, as can be seen already 
in $e^+e^-$ annihilation\cite{LL2000},
the origins of $\Lambda$ are also 
most complicated.
They come not only from the fragmentation 
of $u$, $d$, or $s$-quark, 
but also from the decays 
of many different heavier hyperons,  
such as, $\Sigma^0$, $\Xi^{0,-}$, 
$\Sigma^*(1385)$, and $\Xi^*(1530)$.
Compared with $e^+e^-$ annihilation,
in $\mu^-N\to\mu^-\Lambda X$ 
the contribution from $u$-quark fragmentation 
is much higher in particular for large $x$. 
Since $u$ carries no spin of $\Lambda$ in the 
SU(6) picture and only a small fraction of 
the spin of $\Lambda$ in the DIS picture, 
we expect that the magnitudes of the $P_{\Lambda(L)}$ obtained 
in both models for $\mu^-N\to\mu^-\Lambda X$ 
should be smaller 
than those obtained in $e^+e^-$ annihilation, 
in particular for large $x$.
We expect also that the influences of heavier
hyperon decay are larger. 

Using the event generator {\scriptsize LEPTO}, 
we obtained the $\langle n^a_{\Lambda,f}\rangle$,
$\langle n^b_{\Lambda,H_j}\rangle$,
$\langle n^c_{\Lambda}\rangle$,
and $\langle n^d_{\Lambda}\rangle$
for $\mu^-p\to\mu^-\Lambda X$. 
In Fig.3, we show the different contributions 
in the current fragmentation region 
at $E_{\mu}=500$GeV.
These results are obtained for the kinematic region 
$Q^2>5$(GeV/c)$^2$, $10^{-4}<x<0.2$, and $0.5<y<0.9$.
We choose this kinematic region to ensure a reasonably large $W$
and a reasonably high polarization of 
the fragmenting quark.
We see that 
the contributions from the $s$ quark fragmentation 
are indeed not so dominant as that\cite{LL2000} 
in $e^+e^-\to Z^0\to\Lambda X$.
We also see that in contrast to $s$-quark fragmentation, 
the contributions from the decay of heavier hyperons 
in the case of $u$ or $d$ quark fragmentation 
can even be larger than those directly produced.

Using these results, we obtained 
$P_{\Lambda(L)}$ in $\mu^-p\to\mu^-\Lambda X$, 
as a function of $x_F$ shown in Fig.4
for different combinations of $P^{(l)}_L$ and $P^{(N)}_L$.
We see that except for the case of $P^{(l)}_L=0$ and $P^{(N)}_L=1$,
$P_{\Lambda(L)}$ are positive 
and increases with increasing $x_F$.
Their magnitudes are significantly large,
and that they are different in  
the SU(6) or the DIS picture  
in large $x_F$ region.
We also see that for $P^{(l)}_L=0$ and $P^{(N)}_L=1$, 
$P_{\Lambda(L)}$ is negative
but the magnitude is very small.
This is because in this case the polarizations of the fragmenting quarks
are very small and it is positive for $u$, 
but is negative for $d$ and $s$.
Their contributions to $\Lambda$ polarization partly cancel with each other.
Hence, we see that, to obtain a large longitudinal polarization 
of $\Lambda$, 
it is important to use a polarized $\mu^-$ beam.  

To show the $x$ dependence of $P_{\Lambda(L)}$, 
we calculated also the results 
in the large $x$ region ($0.05<x<0.7$) only.
The obtained results are also given in Fig.4.
We see that they are significantly smaller
than those in the region $10^{-4}<x<0.2$.
This is because for large $x$ the $u$ quarks play the dominant role
and the contribution from $u$ to
$\Lambda$ polarization is small. 

Now we discuss the transversely polarized case.
We calculate the $\Lambda$ polarization under the assumption that 
the same spin transfer mechanism is true
both for longitudinally and transversely polarized cases.
Hence, Eq. (\ref{eq1}) is also true for transversely polarized case,
but the spin transfer factor $t^F_{H_i,f}$ 
can in general be different.
They are determined by the spin structure of hyperon in transversely
polarized case. More precisely,
in the SU(6) picture, $t^F_{H_i,f}$ is the same for transversely
and longitudinally polarized cases.
But, in the DIS picture, 
they are determined by the helicity distribution
$\Delta q_f(x)$ in the longitudinally polarized case,
but by the transversity distribution $\delta q_f(x)$ 
in the transversely polarized case.
Since $\delta q_f(x)$ can in general
be different from $\Delta q_f(x)$, 
the $t^F_{H_i,f}$ in the two cases 
can be different from each other.
Presently, the spin structure in DIS picture
in transversely polarized case is still very unclear,
since no measurement has been done for $\delta q_f(x)$ yet.
In order to see the order of magnitude of the polarization
which we can expect for different hyperon in transversely polarized case,
we made a very rough estimation
by using the same $t^F_{H_i,f}$ as those 
in longitudinally polarized case and $\delta q_f(x)=\Delta q_f(x)$
in calculating $P^{(q)out}_{fT}$ from Eq. (\ref{sub1eq11}).
Under those approximations, the differences between the $\Lambda$
polarization in this case and that in the longitudinally polarized case
of $P_L^{(l)}$
come only from the differences between $D_T(y)$ and $D_L(y)$.

We recall that 
$D_T(y)$ is larger for smaller $y$ (see Fig.1),
so we choose to study in the region $0.1<y<0.7$,
in order to obtain a relatively large transverse polarization of
the fragmenting quark.
We also select the two groups of events in the kinematic region 
$10^{-4}<x<0.2$ and $Q^2>5$ (GeV/c)$^2$, 
and $0.05<x<0.7$ and $Q^2>5$ (GeV/c)$^2$, respectively.
Using the Eqs. (\ref{sub1eq11}) and (\ref{eq1}), 
we obtained $P_{\Lambda (T)}$ in $\mu^-p \to\mu^-\Lambda X$,
as a function of $x_F$ shown in Fig.5a.
We see that $P_{\Lambda (T)}$ is negative 
but its magnitude is relatively small (of the order of 0.02).   

We note that longitudinal $\Lambda$ polarization has been measured \cite{HERMES}
by HERMES Collaboration at HERA in 
$e^+p \to e^+{\Lambda}X$ at 27.5 GeV/c.
However, we can not compare our results shown in Fig.4
with he HERMES data, since the energy in that experiment is rather low.
At that energy, no separation between the fragmentation of
the struck quark and that of the rest of the target proton is 
possible. 
A detailed discussion of this problem is given in Ref. [\ref{LNL}].

\subsubsection{Polarization of other hyperons in $\mu^-p\to\mu^-HX$}

From the results that we obtained in the last subsection, 
we see that, because of the large contributions from 
$u$ and $d$ quark fragmentation in $\mu^-p\to\mu^-\Lambda X$,  
the contributions from the decay of heavier 
hyperons are large and the $\Lambda$ polarization
in transversely polarized case is very small. 
Since the contributions of different flavors to 
the spins of the $J^P=(3/2)^+$ hyperons are 
unknown yet in the DIS picture 
and only models are known to calculate the spin transfer 
in their decay processes, 
there are considerably large theoretical 
uncertainties in the results that we can obtain yet. 
On the other hand, 
as we have already seen \cite{LL2000} in studying 
$e^+e^-\to Z^0\to HX$, 
the origins of other $J^P=(1/2)^+$ hyperons, 
i.e., $\Sigma$ and $\Xi$,  are rather clear.
The decay contributions are much smaller than the corresponding
contributions for $\Lambda$ production.
Hence, the uncertainties in the theoretical calculation in this case 
can be significantly reduced.
We thus continue to study 
the polarization of these $J^P=(1/2)^+$ hyperons 
in $\mu^-N\to\mu^-HX$ in the following.  
We note in particular that,  
in current fragmentation region of $\mu^-p\to\mu^-X$, 
$u$-quark fragmentation plays the dominant role. 
Since the fractional contribution of $u$ quark
to the spin of $\Sigma^+$ is large,
the resulting $\Sigma^+$ 
polarization in the current fragmentation region of 
$\mu^-p\to\mu^-\Sigma^+X$ should be significantly larger than 
that for $\Lambda$.

Furthermore, also because of the large contribution of 
$u$-quark fragmentation,  
the ratio of the production rate of $\Sigma^+$ 
to that of $\Lambda$ in 
the current fragmentation region of $\mu^-p\to\mu^-HX$
should be significantly higher than
from that in $e^+e^-$ annihilation\cite{LL2000},
in particular for large $x_F$ 
where hyperon polarization should be studied.
This can easily be checked using the event generator 
{\scriptsize LEPTO}.
In Fig.6a, we show 
the ratios of the production rate of $\Sigma^+$
to that of $\Lambda$ as function of $x_F$.
We see that the ratio increases 
with increasing $x_F$,
and is already very close to 1 at $x_F=0.8$.
Therefore, the statistics in studying $\Sigma^+$ should not be
much worse than that for $\Lambda$.

In addition to the directly produced, 
there is only one  decay contribution for $\Sigma^+$ 
from $\Sigma^*(1385)$.
In Fig.6b, we show the two
different contributions to $\Sigma^+$ 
in the events where $10^{-4}<x<0.2$.
We see that the decay contribution from heavier hyperon is very small.  
It takes only about $3\%$ in the current fragmentation region.
Hence, the uncertainties in the 
calculations for $\Sigma^+$ are much smaller than those for $\Lambda$
and the study of $\Sigma^+$ polarization
should provide us with a good complementary test 
to different spin transfer pictures.

Using the event generator {\scriptsize LEPTO}, 
we calculated the different average numbers 
$\langle n^a_{\Sigma^+,f}\rangle$,
$\langle n^b_{\Sigma^+,H_j}\rangle$,
$\langle n^c_{\Sigma^+}\rangle$,
and $\langle n^d_{\Sigma^+}\rangle$.
We insert them into Eq. (\ref{eq1}) and obtain 
$P_{\Sigma^+(L)}$ for different combinations of 
$P^{(l)}_L$ and $P^{(N)}_L$ as a function of $x_F$ in Fig.7.
We see that, because of the dominance 
of the contribution from $u$-quark fragmentation, 
and because $u$-quark contributes 
very largely to the spin of $\Sigma^+$, 
the obtained $P_{\Sigma^+(L)}$ is in general 
large and positive in $\mu^- p\to \mu^- \Sigma^+ X$.
We see also that 
the difference between the results
based on the two different pictures
are also larger compared with those for $\Lambda$
and these properties are more significant
for the events with $0.05<x<0.7$.
The results again show that it is important to use a polarized $\mu^-$ beam
to obtain a large longitudinal polarization of $\Sigma^+$.

Similar to $\Lambda$,
we also made a rough calculation of the transverse polarization 
$P_{\Sigma^+(T)}$ of $\Sigma^+$ under the same assumption 
and/or approximations .
The obtained results are shown in Fig.5b.
We see that, compared with those for $\Lambda$,
the magnitude of $P_{\Sigma^+(T)}$ is much larger.
$P_{\Sigma^+(T)}$ has similar properties  as $P_{H_i(L)}$.
It increases 
with increasing $x_F$, and the results obtained in the different pictures
are significantly different from each other in particular at $0.05<x<0.7$.
We see also that $P_{\Sigma^+(T)}$ at $0.05<x<0.7$ is larger
compared with that at $10^{-4}<x<0.2$.
Hence, we expect that the measurement of 
$P_{\Sigma^+(T)}$ 
should provide useful information on
the transverse polarization transfer in high-energy fragmentation processes.

At the same time,
we also make similar calculations for other octet hyperons,
i.e., $\Sigma^-$, $\Xi^{0}$ and $\Xi^{-}$.
The obtained results are shown in Figs.8, 9 and 10. 
We also performed the same calculations
for reactions using neutron target.
Because of the difference caused by the charge factor in
the electromagnetic interactions which make the $u$ (or $d$)
contributions in the case of the neutron target be higher 
(lower) than that of $d$ (or $u$) in the case of proton target,
there is no exact $u$-$d$ exchange symmetry in the results
obtained in the two cases. 
We can not obtain the results for hyperon in the case of neutron
from those in the case of proton by making an interchange of $u$ and $d$.
But the qualitative features can be obtained in this way.
Hence, we expect that measuring $\Sigma^-$ polarization in
$\mu^-n \to \mu^- \Sigma^- X$ can give us useful information
for spin transfer in fragmentation process.


\section{Hyperon polarization
in the current fragmentation region of $\nu_\mu N \to \mu^- H X$}

Compared with $e^+e^-$ annihilation and 
$\mu^-p$ deeply inelastic scattering, 
the deep-inelastic scattering process $\nu_{\mu}N \to \mu^- HX$
has the following two distinct advantages.
First, the magnitude of the polarization of the fragmenting quark is large.
In $\nu_{\mu}N \to \mu^- HX$,
the leading order hard subprocess is $\nu_\mu q \to {\mu}^- q$,
which is a charged current weak interaction
with the exchange of the virtual $W^+$.
According to the standard model for weak interaction,
the charged current weak interaction 
selects only left-handed quarks (right-handed antiquark),
so the polarization of the fragmenting quarks is $P_f^{(q)}=-1$.
Second, there is an automatic flavor separation.
For sufficiently high $Q^2$ and high total hadronic energy $W$, 
the dominant contributions in the current fragmentation region of
$\nu_{\mu} N \to \mu^- HX$ 
come from the fragmentation of the fully polarized $u$ quarks.
Thus, measuring the polarization of the 
hyperon in this reaction
should be able to provide us with a very sensitive test to different models for
the spin transfer in the fragmentation process.

With the aid of {\scriptsize LEPTO} event generator,
we calculated 
$\Lambda$ and $\Sigma^+$ longitudinal polarizations
in the current fragmentation region of 
$\nu_{\mu}p \to \mu^- HX$
at $E_\nu=500$ GeV and $Q^2>5 $(GeV/c)$^2$.
We find out that, the contributions 
from the decay of different heavier hyperons 
to $\Lambda$ are very large.
This can be seen in Fig.11. We see that the decay contribution 
are even larger than those directly produced.
We see also that 
there are significant contributions 
from the decay of $\Lambda_c$ and other charm-baryons.
In particular in the region of $x_F<0.4$,
they are larger than the contributions from the other origins.
Since the decay spin transfer factor $t^D_{H_i,H_j}$ is model dependent
and the polarizations of these heavier hyperons are unclear,
these results imply that there are large theoretical uncertainties
in calculating $P_{\Lambda(L)}$,
in particular from the treatment of $\Lambda_c$ and other charm baryon
decays.
We thus made a very rough estimation for $P_{\Lambda(L)}$ 
using the $t^D_{\Lambda,H_j}$
given in table II in Ref. [\ref{LL2000}] as input.
For $\Lambda_c$, we simply take the fragmentation polarization transfer
from the $c$ quark to $\Lambda^+_c$ is $1$,
and the decay polarization transfer factor for
$\Lambda^+_c\to \Lambda l^+\nu_l$ is also taken as 1,
i.e., $t^F_{\Lambda^+_c,c}=1$ and $t^D_{\Lambda,\Lambda^+_c}=1$.
The obtained results are shown in Fig.12a.
To show the influence of $\Lambda^+_c$ decay,
we show also the results obtained by taking $t^D_{\Lambda,\Lambda^+_c}=0$.
We see that, in the case of $t^D_{\Lambda,\Lambda^+_c}=0$,
$\Lambda$ polarization $P_{\Lambda(L)}$ is very small.
This is because the contributions
from the decay of $\Sigma^*$ and $\Xi^*$ to $\Lambda$ polarization
are negative but other contributions are positive,
and they partly cancel with each other.
We see also that $P_{\Lambda(L)}$
decreases with increasing $x_F$ at $x_F<0.5$
and increases with increasing $x_F$ at $x_F>0.5$.
This is because the contributions
from the decay of $\Sigma^*$ and $\Xi^*$ are
dominant at $x_F<0.5$ and decrease at $x_F>0.5$.
These results obtained in the case of $t^D_{\Lambda,\Lambda^+_c}=1$
differs largely from those in the case of $t^D_{\Lambda,\Lambda^+_c}=0$.
In this case $P_{\Lambda(L)}$ is still negative,
but the magnitude of $P_{\Lambda(L)}$ is much larger than 
that in the case of $t^D_{\Lambda,\Lambda^+_c}=0$. 
The decrease with increasing $x_F$ at $x_F<0.4$
and the increase with increasing $x_F$ at $x_F>0.4$
is due to the contributions from $\Lambda^+_c$ decay 
are dominant at $x_F<0.4$ and decrease at $x_F>0.4$.

In contrast to $\Lambda$, from Fig.12b and 12d, we see that 
the contributions from the decay of heavier strange hyperons 
and charm-baryons to $\Sigma^+$ are very small.
The contributions from the decay of $\Sigma^*$
containing the fragmenting $u$ quark only take about $1\%$,
and that those from the decay of $\Lambda_c$ and other charm-baryons
take only about $8\%$.
Therefore the theoretical uncertainties 
in calculating $\Sigma^+$ polarization are small. 
Furthermore,
because of the dominance of the $u$ quark fragmentation,
we expected that the $\Sigma^+$ production rate should be relatively high.
This can easily be seen in Fig.12 or Fig. 13
where we show the ratio of the production rate of $\Lambda$ to 
that of $\Sigma^+$.
The result shows that the ratio increases with the increasing $x_F$,
and reaches about 1.6 at $x_F=0.8$.
Hence, the statistics for $\Sigma^+$ is not less than that for $\Lambda$
in the current fragmentation region of 
$\nu_{\mu} p \to \mu^- \Lambda /\Sigma^++X$.

Inserting these results into Eq.(\ref{eq1})
we obtain the $\Sigma^+$ polarization shown in Fig.12b,
where the contributions from the decay of charm-baryons
are not taken into account.
We see that $\Sigma^+$ polarization is negative 
and its magnitude is very large and 
increases with increasing $x_F$,
and the difference between the results 
based on the two different pictures
is quite significant.
Hence measuring $P_{\Sigma^+(L)}$
in the current fragmentation region of  
$\nu_{\mu} p \to \mu^- \Sigma^+X$
should be very validity to distinguish the different pictures
for spin transfer in high-energy fragmentation processes.

\section{Summary}

We calculated the polarizations
for different octet hyperons
produced in the current fragmentation regions
of the deeply inelastic lepton-nucleon scatterings
$\mu^-N \to \mu^- HX$ and
$\nu_{\mu} N \to \mu^- HX$ for sufficiently large $Q^2$ and $W^2$
using different models for spin transfer
in high-energy fragmentation processes.
We discussed in detail
the contributions from different sources including those
from the decay of different heavier hyperons.
Our results show that measurements
of these hyperon polarizations
should provide useful information
to distinguish between different models
in particular the SU(6) 
and the DIS pictures used frequently in the literature.
We found in particular that
decay contributions to $\Lambda$ are very important 
in $\mu^-N\to\mu^-\Lambda X$ 
and $\nu N\to\mu^-\Lambda X$ 
and have to be taken into account in 
calculating the $\Lambda$ polarizations 
in these reactions.  
We found also that measuring 
the polarizations of $\Sigma^+$ 
in $\mu^-p\to \mu^-\Sigma^+X$
or $\nu_\mu p\to \mu^-\Sigma^+X$
can give a much better test
to the validity of the different models
than that obtained by measuring $\Lambda$ 
in the corresponding processes not only because 
the magnitudes of $P_{\Sigma^+}$ and the differences 
between different models are much larger than 
that of $P_\Lambda$ but also because 
the theoretical uncertainties are much smaller 
in calculating the former.
Measuring $P_{\Sigma^+(T)}$ 
in transversely polarized case should 
be able to give an useful check 
whether in high-energy fragmentation process
the polarization transfer mechanisms 
are the same for 
both the longitudinally polarized and transversely polarized cases.

\vskip 1.0cm

We thank Li Shi-yuan, Xie Qu-bing
and other members in the theoretical particle physics group of
Shandong University for helpful discussions.
This work was supported in part by the National Science Foundation
of China (NSFC) and the Education Ministry of China.

\newpage
\begin{figure}[ht]
\psfig{file=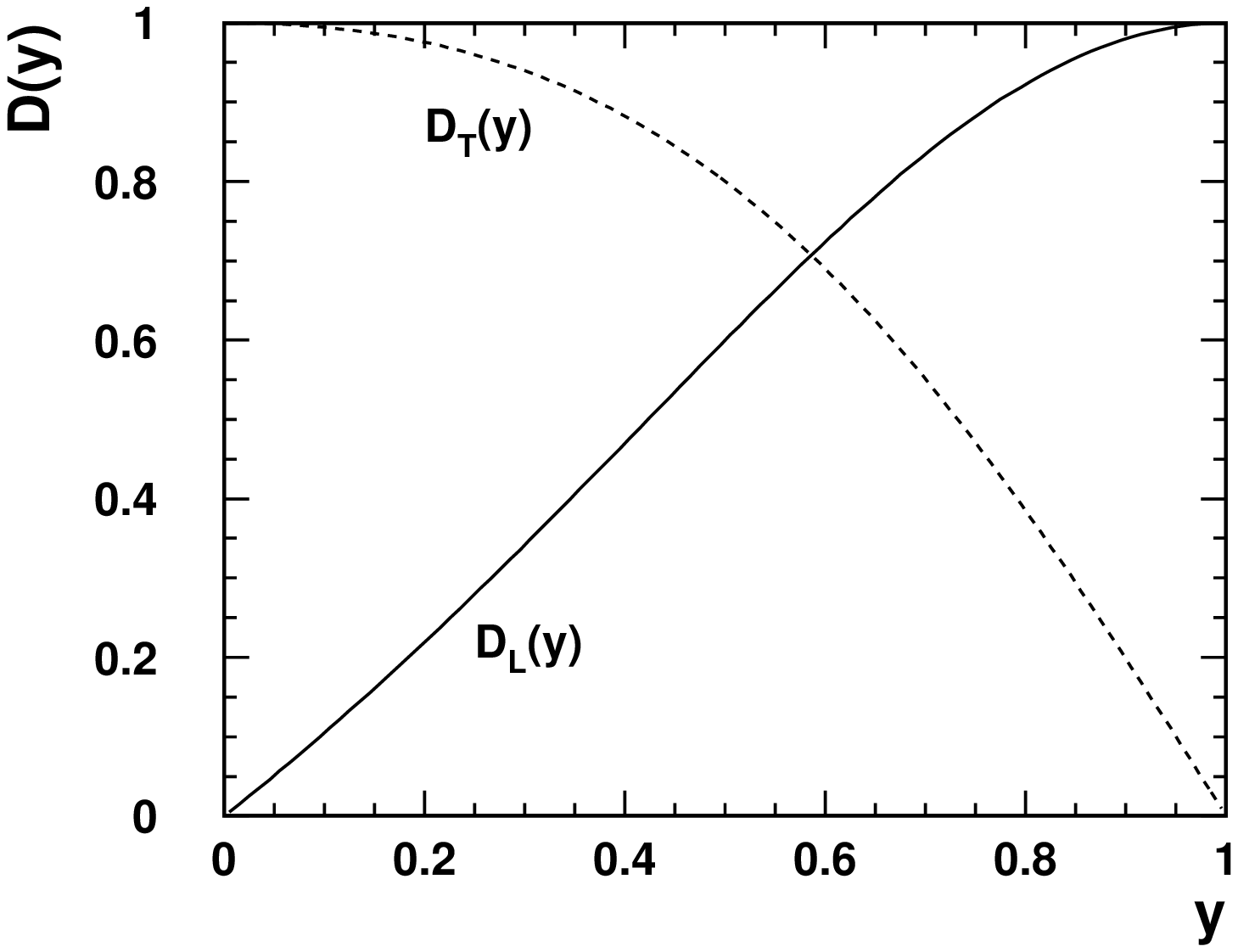,width=8cm}
\caption{Depolarization factors $D_L(y)$ and $D_T(y)$ 
as functions of $y$ in $\mu^-q\to\mu^-q$ for the 
scattered quark.}
\label{fig1}
\end{figure}

\begin{figure}[ht]
\psfig{file=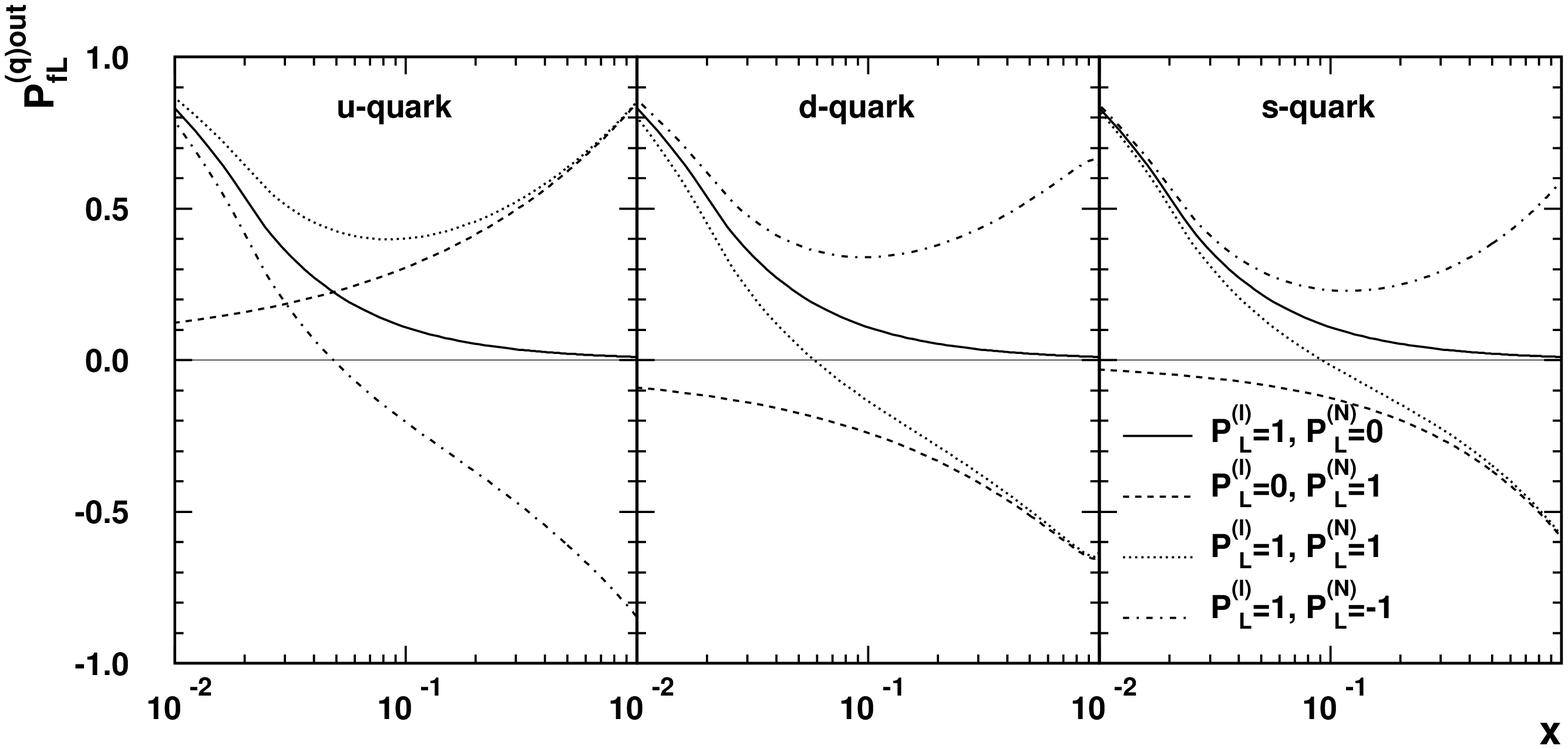,width=12cm}
\caption{The $x$ dependences of the polarization $P_{fL}^{(q)out}$ of
the outgoing scattered $q_f$ in $\mu^-p\to \mu^-X$ 
for different combinations of $P^{(l)}_L$ and $P^{(N)}_L$
at $Q^2=10$ (GeV/c)$^2$ and $E_{\mu}$=500 GeV.}
\label{fig2}
\end{figure}

\begin{figure}[ht]
\psfig{file=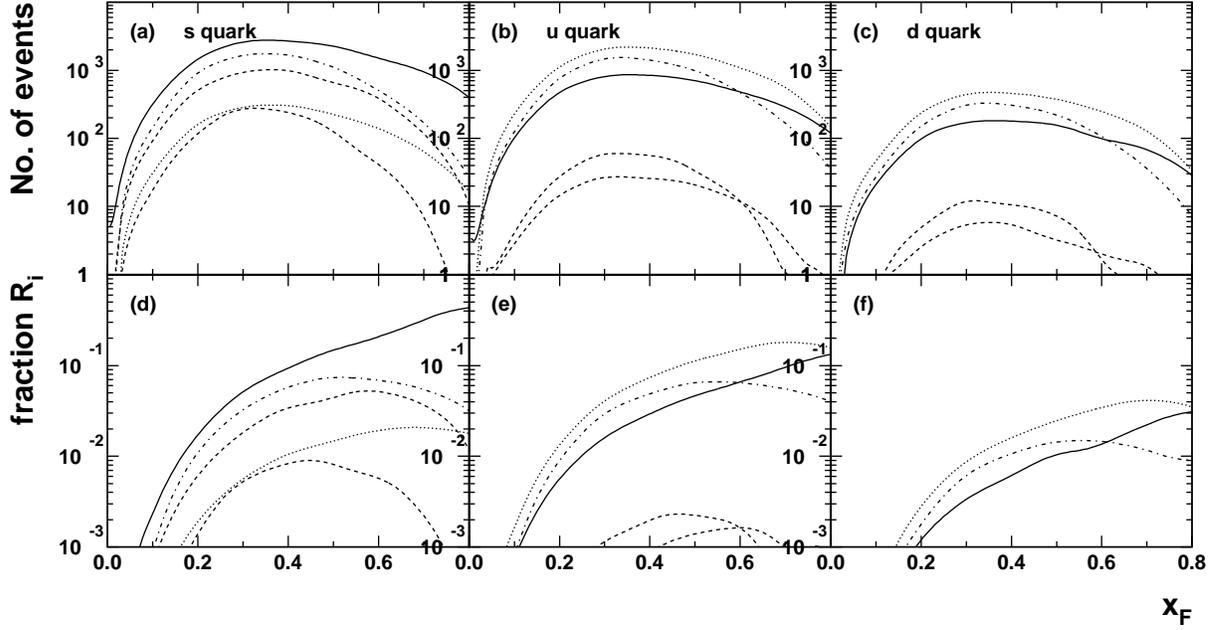,height=10cm}
\caption{Different contributions to $\Lambda$ in events
originating from the $s$, $u$, or $d$ quark fragmentation
as functions of $x_F=2p^*_z/W$ in $\mu^-p \to \mu^- \Lambda X$
at $E_\mu=500$GeV.
In (a)-(c), we see the five types of contributions 
from $s$, $u$, or $d$ quark respectively.
The solid line denotes those which are directly produced 
and contain the fragmenting quark;
the {\it upper} dashed, {\it lower} dashed,
dotted, and dash-dotted lines respectively 
denote those from the decay of 
$\Xi$, $\Xi^*$, $\Sigma^0$, and $\Sigma^*$
which contain the fragmenting quark.
In (d)-(f), we see the fractions $R_i$ 
which are the ratios of the 
corresponding contributions
to the sum of all different contributions.}
\label{fig3}
\end{figure}

\begin{figure}[ht]
\psfig{file=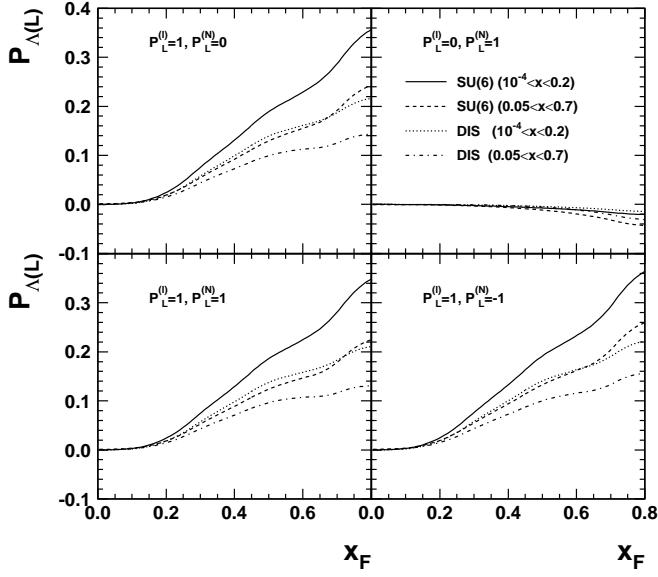,width=10cm}
\caption{Longitudinal polarization $P_{\Lambda(L)}$
of $\Lambda$ as a function of $x_F$ for the different 
combinations of $P^{(l)}_L$ and $P^{(N)}_L$.
The solid and dotted lines are respectively the results 
based on the SU(6) and the DIS pictures for the events
with $10^{-4}<x<0.2$;
the dashed and dash-dotted lines are the corresponding results 
for the events with $0.05<x<0.7$.}
\label{fig4}
\end{figure}

\begin{figure}[ht]
\psfig{file=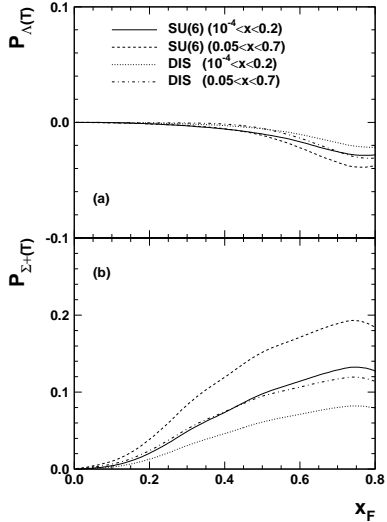,width=6cm}
\caption{Transverse polarization of 
$\Lambda$ and $\Sigma^+$, $P_{\Lambda(T)}$
and $P_{\Sigma^+(T)}$, as functions of $x_F$.}
\label{fig5}
\end{figure}

\begin{figure}[ht]
\psfig{file=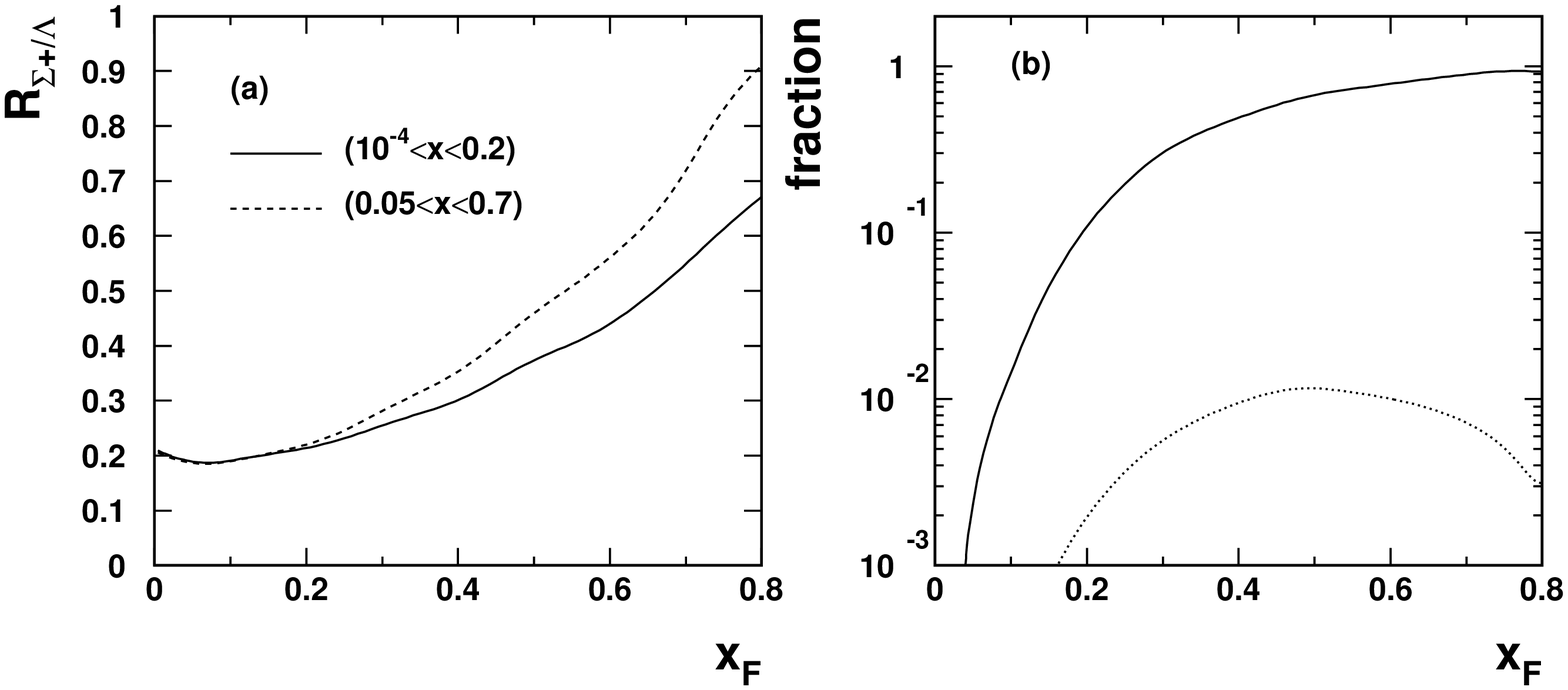,width=13cm}
\caption{(a): Ratio of the production rate of $\Sigma^+$ to
that of $\Lambda$ as a function of $x_F$.
The solid and dashed lines are respectively 
the results for the events where $10^{-4}<x<0.2$
and the events where $0.05<x<0.7$.
(b): Different contributions to $\Sigma^+$ in 
$\mu^-p\to\mu^-\Sigma^+X$ at $E_\mu=500$ GeV. 
The solid and dashed lines are respectively 
the contributions
which are directly produced and contain the fragmenting quark
and those which are originate from 
the decay of polarized heavier hyperons.}
\label{fig6}
\end{figure}

\begin{figure}[ht]
\psfig{file=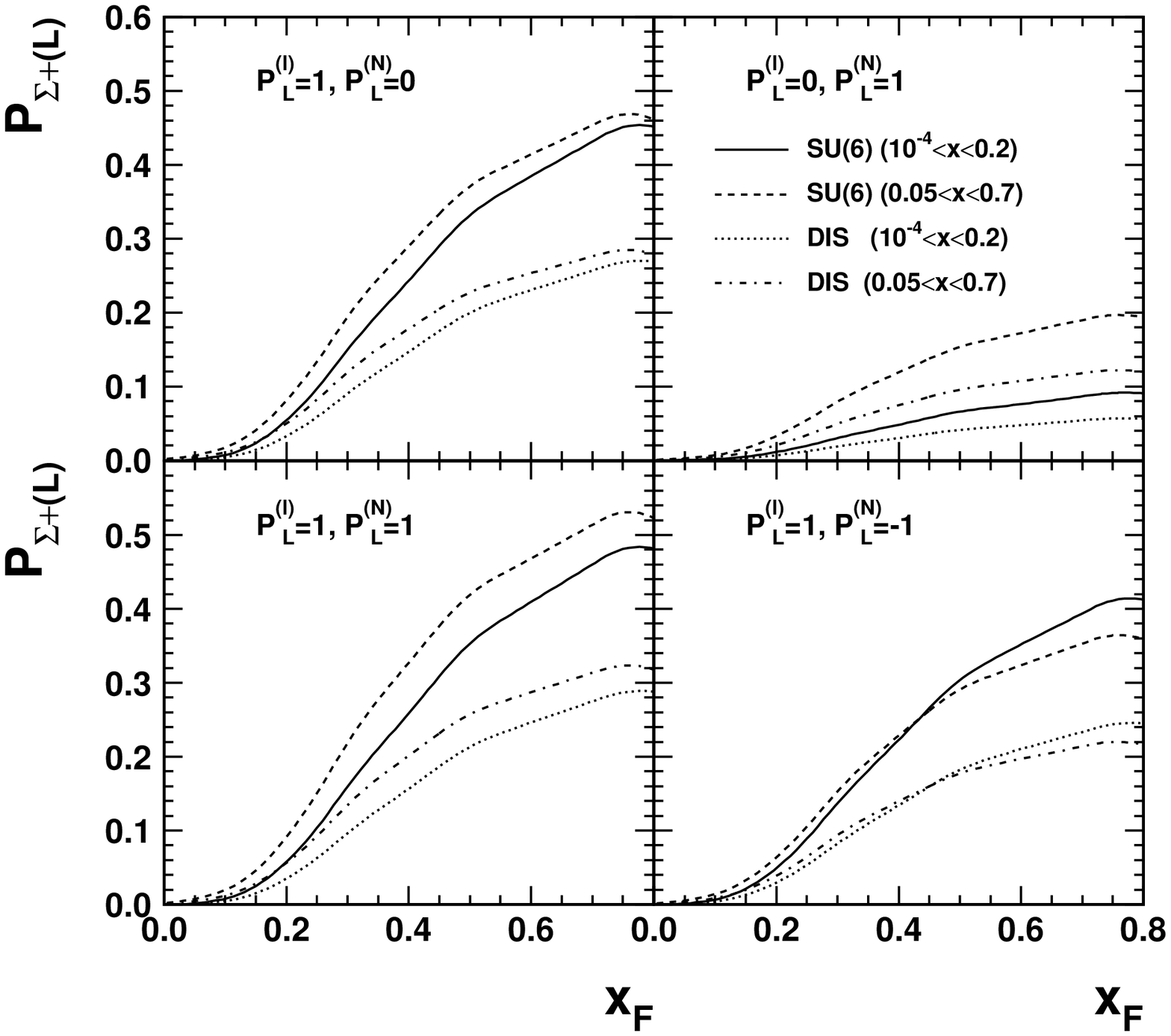,width=10cm}
\caption{ Longitudinal polarization $P_{\Sigma^+(L)}$
of $\Sigma^+$ in the current fragmentation region 
of $\mu^p\to\mu^-\Sigma^+X$ as a function of $x_F$.}
\label{fig7}
\end{figure}

\begin{figure}[ht]
\psfig{file=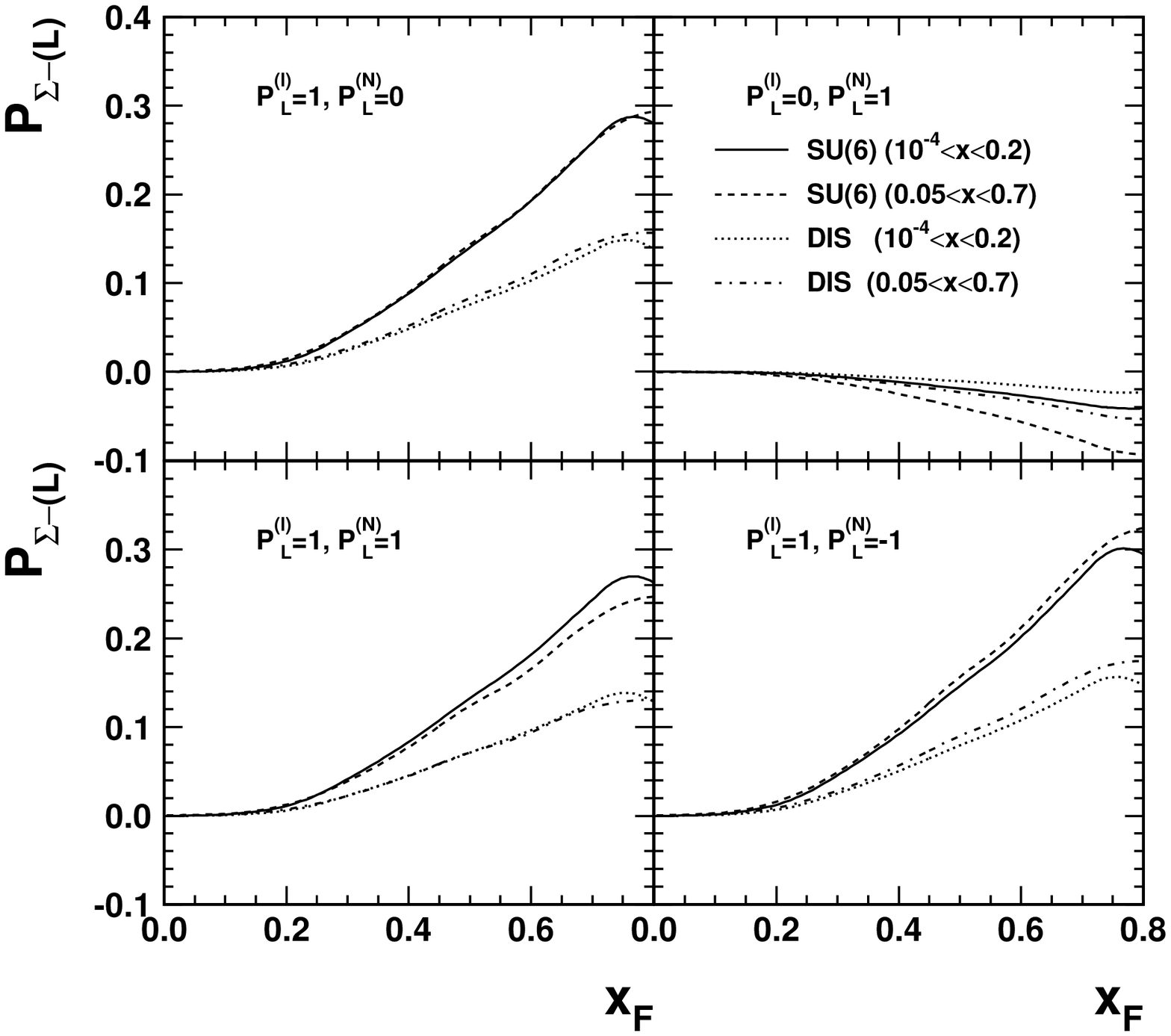,width=10cm}
\caption{ Longitudinal polarization $P_{\Sigma^-(L)}$
of $\Sigma^-$ in the current fragmentation region of 
$\mu^-p\to\mu^-\Sigma^-X$ as a function of $x_F$.}
\label{fig8}
\end{figure}

\begin{figure}[ht]
\psfig{file=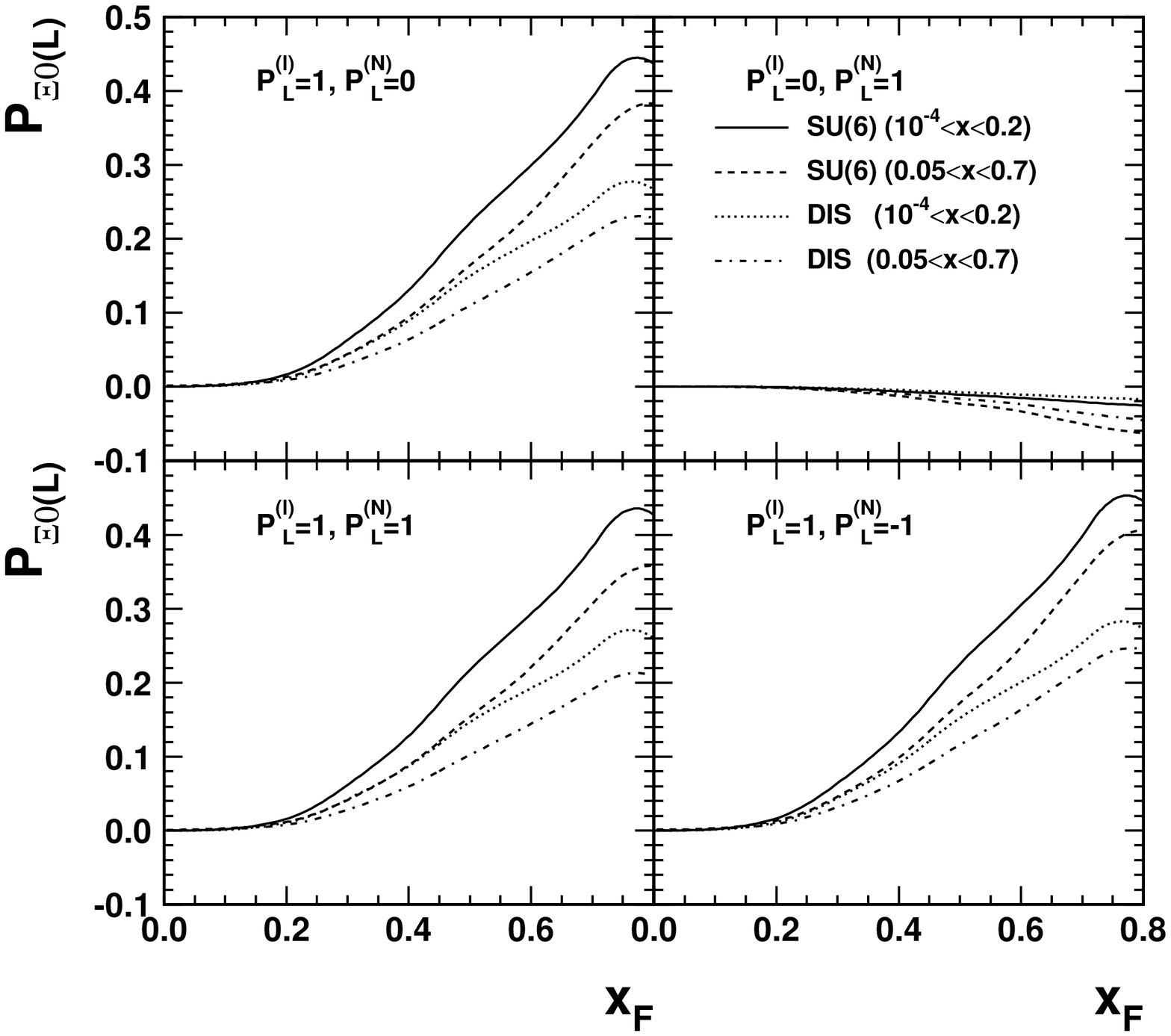,width=10cm}
\caption{Longitudinal polarization 
$P_{\Xi^0(L)}$ of $\Xi^0$ in 
the current fragmentation region of 
$\mu^-p\to\mu^-\Xi^0X$ as a function of $x_F$.}
\label{fig9}
\end{figure}

\begin{figure}[ht]
\psfig{file=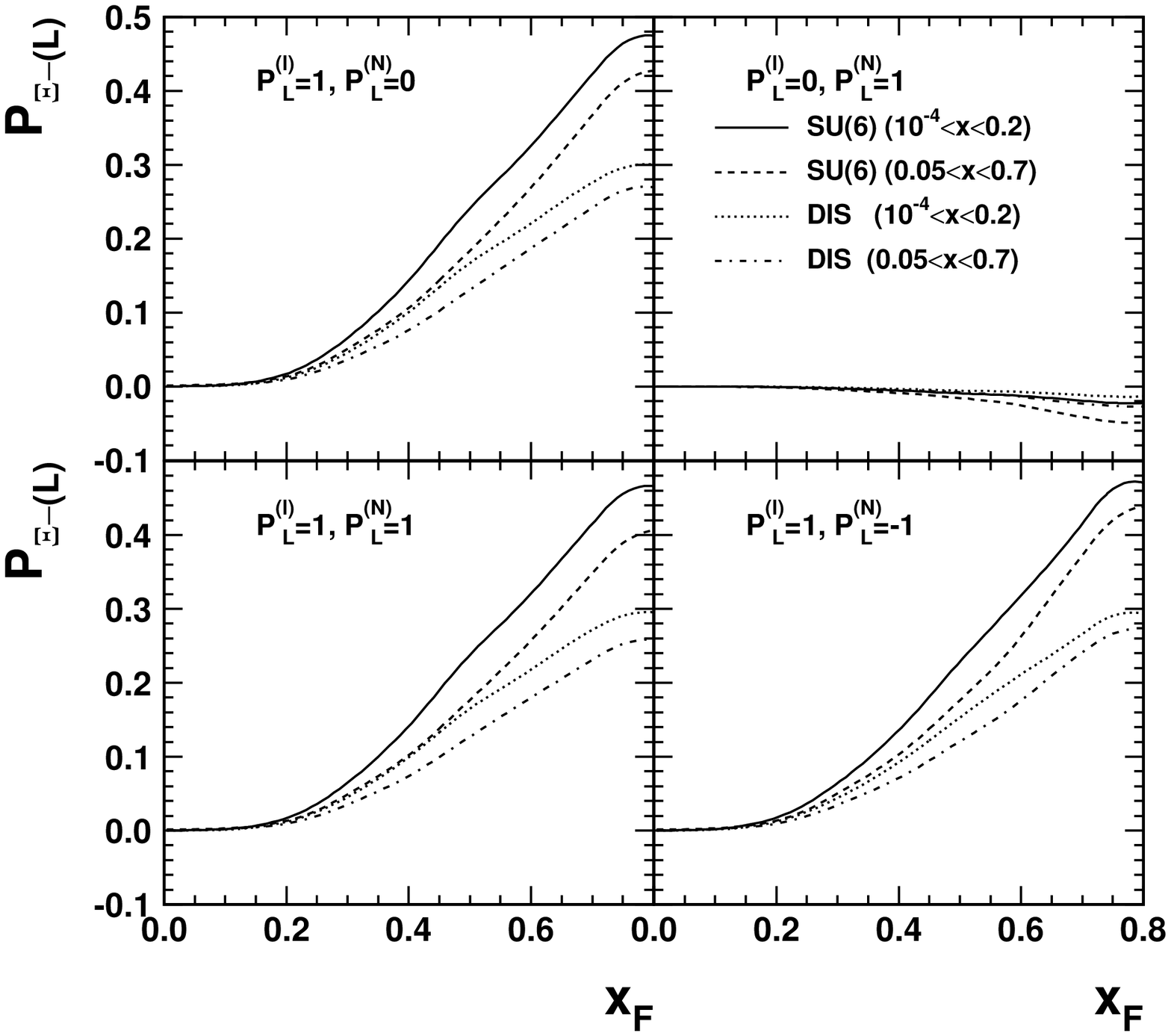,width=10cm}
\caption{Longitudinal polarization 
$P_{\Xi^-(L)}$ of $\Xi^-$ in 
the current fragmentation region of 
$\mu^-p\to\mu^-\Xi^-X$ as a function of $x_F$.}
\label{fig10}
\end{figure}

\newpage
\begin{figure}[t]
\psfig{file=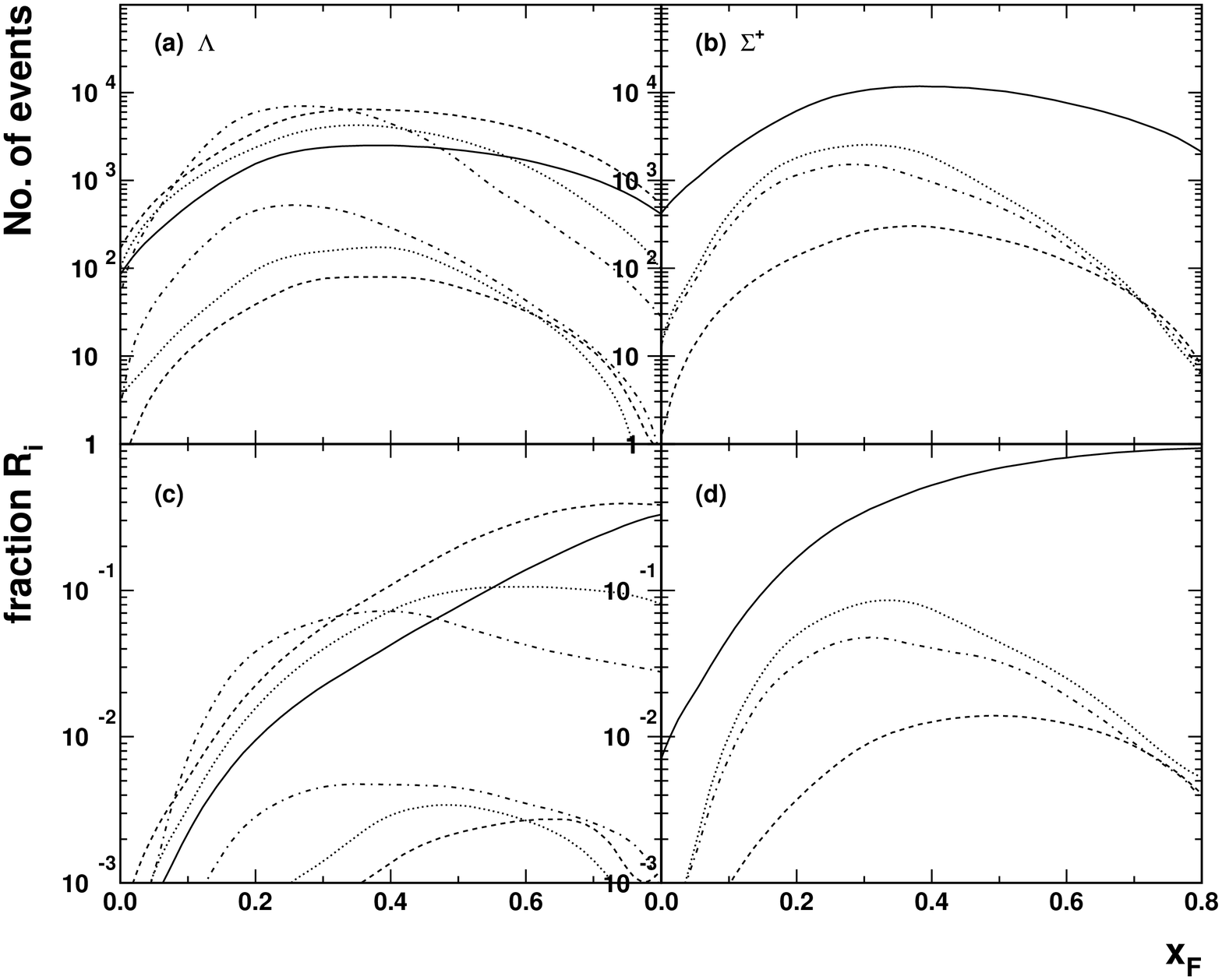,width=10cm}
\caption{Different contributions to $\Lambda$ or $\Sigma^+$
in the fragmentation of $u$, and $c$ quarks
as functions of $x_F$ in the current fragmentation region of
$\nu_{\mu}p\to\mu^-HX$ at $E_\nu=500$ GeV.
In (a), we see the seven types of contributions to $\Lambda$.
The solid, {\it upper} dashed, {\it upper} dotted,
{\it lower} dashed, and {\it lower} dotted lines 
denote those which are directly 
produced or from the decay of
$\Sigma^0$, $\Sigma^*$, $\Xi$, and $\Xi^*$
that contain the fragmenting $u$ quark;
the {\it upper} dash-dotted and {\it lower} dash-dotted lines
denote those from the decay of $\Lambda_c^+$ and other charm-baryons.
In (b), we see the four types of contributions to $\Sigma^+$
in the fragmenting $u$ and $c$ quark events.
Here, the solid and dashed lines denote 
those which are directly produced
and contain the fragmenting $u$ quark and 
those from the decay of $\Sigma^*$ that  
contain the fragmenting $u$ quark respectively;
the dotted and dash-dotted lines
denote those from the decay of $\Lambda_c^+$ and other charm-baryons.
In (c) and (d), we see the fractions $R_i$
which are the ratios of the corresponding contributions
to the sums of all these different contributions 
for $\Lambda$ and $\Sigma^+$ respectively.}
\label{fig11}
\end{figure}

\begin{figure}[ht]
\psfig{file=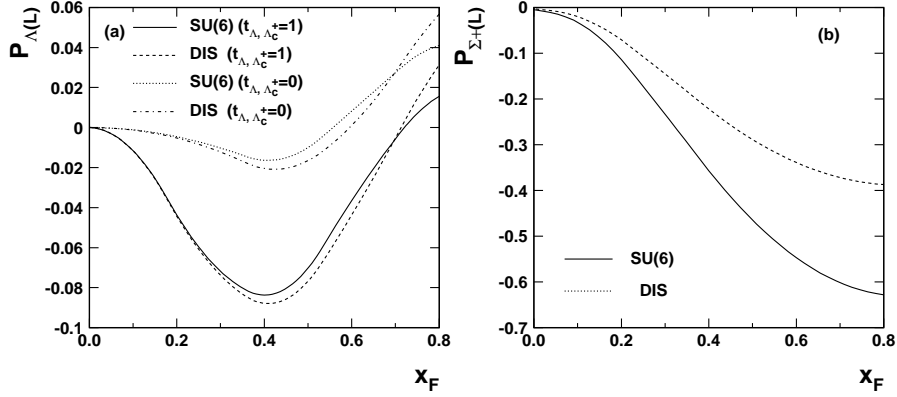,width=13cm}
\caption{Longitudinal polarizations of $\Lambda$ and $\Sigma^+$ 
in the current fragmentation region of 
$\nu_\mu p\to \mu^- HX$.
In (a),
the solid and dashed lines are respectively the results
based on the SU(6) and DIS pictures 
at $t^D_{\Lambda,\Lambda^+_c}=1$;
the dotted and dash-dotted lines are respectively the results 
based on the SU(6) and DIS pictures 
at $t^D_{\Lambda,\Lambda^+_c}=0$.
In (b), the solid and dashed lines are respectively the results obtained
based on the SU(6) and DIS pictures,
where the decay of charm-baryons are not taken into account.}
\label{fig12}
\end{figure}

\begin{figure}[ht]
\psfig{file=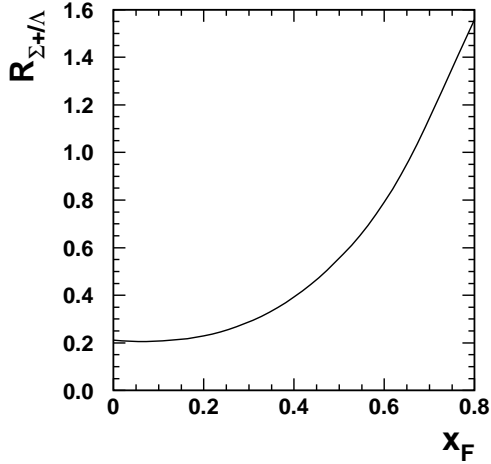,width=8cm}
\caption{The ratio of the production rate of $\Sigma^+$ to
that of $\Lambda$ as a function of $x_F$
in the current fragmentation region of $\nu_\mu p\to\mu^- HX$.}
\label{fig13}
\end{figure}

\newpage
Figure captions

\vskip 0.3cm\noindent
Fig.1: Depolarization factors $D_L(y)$ and $D_T(y)$ 
as functions of $y$ in $\mu^-q\to\mu^-q$ for the 
scattered quark.

\vskip 0.3cm\noindent
Fig.2: The $x$ dependences of the polarization $P_{fL}^{(q)out}$ of
the outgoing scattered $q_f$ in $\mu^-p\to \mu^-X$ 
for different combinations of $P^{(l)}_L$ and $P^{(N)}_L$
at $Q^2=10$ (GeV/c)$^2$ and $E_{\mu}$=500 GeV.

\vskip 0.3cm\noindent
Fig.3: Different contributions to $\Lambda$ in events
originating from the $s$, $u$, or $d$ quark fragmentation
as functions of $x_F=2p^*_z/W$ in $\mu^-p \to \mu^- \Lambda X$
at $E_\mu=500$GeV.
In (a)-(c), we see the five types of contributions 
from $s$, $u$, or $d$ quark respectively.
The solid line denotes those which are directly produced 
and contain the fragmenting quark;
the {\it upper} dashed, {\it lower} dashed,
dotted, and dash-dotted lines respectively 
denote those from the decay of 
$\Xi$, $\Xi^*$, $\Sigma^0$, and $\Sigma^*$
which contain the fragmenting quark.
In (d)-(f), we see the fractions $R_i$ 
which are the ratios of the 
corresponding contributions
to the sum of all different contributions.

\vskip 0.3cm\noindent
Fig.4: Longitudinal polarization $P_{\Lambda(L)}$
of $\Lambda$ as a function of $x_F$ for the different 
combinations of $P^{(l)}_L$ and $P^{(N)}_L$.
The solid and dotted lines are respectively the results 
based on the SU(6) and the DIS pictures for the events
with $10^{-4}<x<0.2$;
the dashed and dash-dotted lines are the corresponding results 
for the events with $0.05<x<0.7$.

\vskip 0.3cm\noindent
Fig.5: Transverse polarization of 
$\Lambda$ and $\Sigma^+$, $P_{\Lambda(T)}$
and $P_{\Sigma^+(T)}$, as functions of $x_F$.

\vskip 0.3cm\noindent
Fig.6:(a): Ratio of the production rate of $\Sigma^+$ to
that of $\Lambda$ as a function of $x_F$.
The solid and dashed lines are respectively 
the results for the events where $10^{-4}<x<0.2$
and the events where $0.05<x<0.7$.
(b): Different contributions to $\Sigma^+$ in 
$\mu^-p\to\mu^-\Sigma^+X$ at $E_\mu=500$ GeV. 
The solid and dashed lines are respectively 
the contributions
which are directly produced and contain the fragmenting quark
and those which are originate from 
the decay of polarized heavier hyperons.

\vskip 0.3cm\noindent
Fig.7: Longitudinal polarization $P_{\Sigma^+(L)}$
of $\Sigma^+$ in the current fragmentation region 
of $\mu^p\to\mu^-\Sigma^+X$ as a function of $x_F$.

\vskip 0.3cm\noindent
Fig.8: Longitudinal polarization $P_{\Sigma^-(L)}$
of $\Sigma^-$ in the current fragmentation region of 
$\mu^-p\to\mu^-\Sigma^-X$ as a function of $x_F$.

\vskip 0.3cm\noindent
Fig.9: Longitudinal polarization $P_{\Xi^0(L)}$ of $\Xi^0$ in 
the current fragmentation region of 
$\mu^-p\to\mu^-\Xi^0X$ as a function of $x_F$.

\vskip 0.3cm\noindent
Fig.10: Longitudinal polarization 
$P_{\Xi^-(L)}$ of $\Xi^-$ in 
the current fragmentation region of 
$\mu^-p\to\mu^-\Xi^-X$ as a function of $x_F$.

\vskip 0.3cm\noindent
Fig.11: Different contributions to $\Lambda$ or $\Sigma^+$
in the fragmentation of $u$, and $c$ quarks
as functions of $x_F$ in the current fragmentation region of
$\nu_{\mu}p\to\mu^-HX$ at $E_\nu=500$ GeV.
In (a), we see the seven types of contributions to $\Lambda$.
The solid, {\it upper} dashed, {\it upper} dotted,
{\it lower} dashed, and {\it lower} dotted lines 
denote those which are directly 
produced or from the decay of
$\Sigma^0$, $\Sigma^*$, $\Xi$, and $\Xi^*$
that contain the fragmenting $u$ quark;
the {\it upper} dash-dotted and {\it lower} dash-dotted lines
denote those from the decay of $\Lambda_c^+$ and other charm-baryons.
In (b), we see the four types of contributions to $\Sigma^+$
in the fragmenting $u$ and $c$ quark events.
Here, the solid and dashed lines denote 
those which are directly produced
and contain the fragmenting $u$ quark and 
those from the decay of $\Sigma^*$ that  
contain the fragmenting $u$ quark respectively;
the dotted and dash-dotted lines
denote those from the decay of $\Lambda_c^+$ and other charm-baryons.
In (c) and (d), we see the fractions $R_i$
which are the ratios of the corresponding contributions
to the sums of all these different contributions 
for $\Lambda$ and $\Sigma^+$ respectively.

\vskip 0.3cm\noindent
Fig.12: Longitudinal polarizations of $\Lambda$ and $\Sigma^+$ 
in the current fragmentation region of 
$\nu_\mu p\to \mu^- HX$.
In (a), the solid and dashed lines are respectively the results
based on the SU(6) and DIS pictures 
at $t^D_{\Lambda,\Lambda^+_c}=1$;
the dotted and dash-dotted lines are respectively the results 
based on the SU(6) and DIS pictures 
at $t^D_{\Lambda,\Lambda^+_c}=0$.
In (b), the solid and dashed lines are respectively the results obtained
based on the SU(6) and DIS pictures,
where the decay of charm-baryons are not taken into account.

\vskip 0.3cm\noindent
Fig.13: The ratio of the production rate of $\Sigma^+$ to
that of $\Lambda$ as a function of $x_F$
in the current fragmentation region of $\nu_\mu p\to\mu^- HX$.


\begin {thebibliography}{99}
\bibitem{ref1} 
We note that at HERMES or NOMAD energies, 
the total hadronic squared, $W^2$, 
is typically of the order of 10 (GeV)$^2$.
Hence, W is only of several GeV.
In this case, no separation of the fragments of the scattered 
quark and those of the rest of the nucleon is possible. 
Hence, the contamination from usual target fragments 
to the current fragmentation is very high.
\end{thebibliography}
\end{document}